\documentstyle[prl,aps]{revtex}
\twocolumn
\begin{document}
\twocolumn[\hsize\textwidth\columnwidth\hsize\csname @twocolumnfalse\endcsname

\title{Disorder effects in electronic structure of substituted 
       transition metal compounds}
\author{D.D. Sarma\cite{jnc}, A. Chainani\cite{add} and S.R. Krishnakumar}
\address{Solid State and Structural Chemistry Unit, Indian Institute
of Science, Bangalore 560 012, INDIA}
\author{E. Vescovo, C. Carbone and W. Eberhardt}
\address{Institut f\"ur Festk\"orperforschung, Forschungszentrum
J\"ulich, D-52425 J\"ulich, GERMANY}
\author{O. Rader, Ch. Jung, Ch. Hellwig and W. Gudat}
\address{BESSY, Lentzeallee 100, D-14195 Berlin, GERMANY}
\author{H. Srikanth and A.K. Raychaudhuri}
\address{Department of Physics, Indian Institute of Science,
Bangalore 560 012, INDIA}
\date{\today}
\maketitle
\begin{abstract}

Investigating LaNi$_{1-x}M_x$O$_3$ ($M$ = Mn and Fe), we identify
a characteristic evolution of the spectral function with increasing 
disorder in presence of strong interaction effects across the 
metal-insulator transition. We discuss these results {\it vis-a-vis} 
existing theories of electronic structure in simultaneous presence 
of disorder and interaction.

\end{abstract}
]

The last few years have seen a spectacular resurgence of interest in
investigating the phenomena of metal-insulator transitions (MIT) in
substituted transition metal oxides (TMO). MIT which are driven by 
electron interaction effects as in the Mott-Hubbard
picture\cite{mott,hubbard} 
or by disorder effects causing an Anderson 
transition\cite{anderson}, have led to distinctly
different paradigms. 
However, experimental realisations
of MIT in substituted TMO's show signatures of both these effects. 
A few theoretical attempts\cite{AA,CG,TVR,EPL,GabiPRL} 
have been made to study the electronic structure in 
simultaneous presence of interaction and disorder. However, 
these do not provide a unifying description of the 
ground state electronic structure and the MIT.
Briefly summarizing, Altshuler-Aronov~(AA) theory for disordered
metals\cite{AA} 
predicts a characterisitc $|E-E_F|^{1/2}$ cusp at $E_F$
progressively depleting the density of states (DOS) at the
Fermi energy, $N(E_F)$, on approaching the
MIT boundary from the metallic side.
On the other hand, a disorder driven insulator would open 
up a soft Coulomb gap\cite{CG}
characterised by a $(E-E_F)^2$ dependence of DOS near 
$E_F$ in presence of long range
interactions. The AA theory, restricting itself to a narrow energy 
scale within the
metallic state, or the Coulomb gap
 insulating phase, does not provide a description for either 
the evolution of the electronic structure across the transition, 
or the electronic
structure over a larger energy scale
($\sim U$ or $W$). 
Treating diagonal disorders in a Hubbard-like model
within the dynamical mean field theory\cite{RMP}, it has been
predicted\cite{EPL,priv} that the electronic structure of a substituted
TMO resembles an additive average of the electronic structures of the
two stoichiometric end-members on a wide energy scale. While this is
the only prediction available for the evolution of electronic
structures over a wide energy scale for disordered substituted TMO,
it suggests the formation of a {\it narrow resonance peak}
at $E_F$ within the metallic phase
in 
contrast to the AA theory. 
Additionally, while these conflicting scenarios relate only
to ground state electronic structures, experimentally various
interesting cross-overs of transport properties are often
observed as a function of temperature, suggesting changes in the
electronic structure over a low energy scale with changing
temperature\cite{NiS2}.

In most cases of substituted TMO, the MIT is achieved by altering the
composition via substitution chemistry, thereby tuning the charge
carrier density by doping. While this way of doping charge carriers
{\it invariably} introduces disorder in the system, this aspect is
often ignored. This has
led to the wide spread use of a homogeneous model like the
Hubbard model which includes only interaction effects, to describe the
evolution of electronic structures in these systems. However, in many
cases there are obvious experimental signatures not understood
within Hubbard-like models; 
for example, at very low temperatures, La$_{2-x}$Sr$_x$CuO$_4$ with 
$x$=0.02 shows 
Coulomb gap behavior which changes to variable range hopping (VRH)
for x = 0.05\cite{rosen}. The need to dope macroscopic charge carrier
concentrations (often upto 0.2 or even more per transition metal
site) to drive the insulator-to-metal transition is also 
beyond the scope
of any homogeneous model. However, disorder induced changes in the 
electronic structure are often difficult to probe in presence of
chemical substitutions, since 
such substitutions result in large changes 
in the electronic
structure due to the doping of new (electron or hole) states; 
these tend to dominate over every other subtle change 
in the electronic structure. Additionally, doping 
induced structural transitions can make it impossible to
extricate the effects of disorder {\it vis-a-vis} correlations.

In order to study the evolution of electronic structure of a
correlated system exhibiting a MIT due to static disorder but in
absence of (i) charge
carrier alteration in the form of doped states 
as well as (ii) a structural instability, we have carried out a
detailed spectroscopic investigation of the isostructural 
series, 
LaNi$_{1-x}M_x$O$_3$ ($M$ = Mn, Fe or Co). While
LaNiO$_3$ is a correlated metallic oxide\cite{corr}, La$M$O$_3$ with
$M$ = Mn, Fe and Co are insulating and the solid solutions 
LaNi$_{1-x}M_x$O$_3$ exhibit\cite{MIT} MIT with increasing $x$.  In
early work, this transition was discussed essentially in terms of
percolation models, since the end members of the solid solution have
differing ground state electronic properties.  However percolation
theory cannot obviously account for very different values of the
critical concentration, $x_c$ with changing $M$ : $x_c$(Mn) $\approx
0.1$, $x_c$(Fe) $\approx 0.2$ and $x_c$(Co) $\approx 0.65$, 
for the same structure, with lattice 
parameters remaining more or less unaltered.  Moreover, this
is a case of homovalent substitution (Ni$^{3+}$ by $M^{3+}$) as
against heterovalent substitution ({\it e.g.} La$^{3+}$ by Sr$^{2+}$,
which dopes hole states).  This makes the system ideally suited for
investigating possible manifestations of disorder effects within the 
strongly interacting transition metal oxide system.  We show
that indeed there is a pronounced cusp in the DOS at low energy
scales (within 200~meV of $E_F$) in contrast to the resonance
expected within the Hubbard-like model. Interestingly however,
changes in the electronic structure over a wider energy scale is in
agreement with this model, suggesting
that the existing theories are only partially successful in
describing the evolution of the electronic structure with changing
disorder in such strongly correlated systems. Additionally, spectral 
functions exhibit changes that correlate with the cross-over of the 
transport behavior as a function of temperature.

The sample preparation and characterization have been
described earlier \cite{sample}.  Photoemission experiments
were carried out at SX-700-II and 3M-NIM-I beamlines of BESSY using a
high-resolution VG ESCALAB system with a resolution of
about 100 meV. Bremsstrahlung isochromat spectra (BIS) were 
recorded in a 
VSW spectrometer with a resolution of 
about 0.7~eV. The samples were maintained at 100~K 
to avoid surface degradation and intermittently scraped {\it
in-situ} to obtain a clean surface.  Point-contact 
conductance spectra were obtained at 4.2~K, using electrochemically
etched gold tips.

We show BIS of LaNi$_{1-x}$Mn$_x$O$_3$ 
in Fig.~1 which is dominated by transition metal 3$d$
states over the first 3~eV with minimal overlap from other states.
LaNiO$_3$ spectrum clearly shows a large
intensity, while LaMnO$_3$ exhibits negligible intensity at $E_F$,
consistent with the metallic and insulating properties respectively. 
It turns out that the spectra of the
intermediate compositions can be well described by a weighted
average of the spectra of the end-members (LaNiO$_3$ and LaMnO$_3$), 
as shown by solid 
lines superimposed on the experimental data.
This remarkable agreement over a wide energy scale is
consistent with the dynamical mean-field result
of a Hubbard-like model with diagonal disorder\cite{EPL,priv}.
While similar conclusions can also be drawn on the basis of oxygen
$K$-edge x-ray absorption spectra of all the three series,
LaNi$_{1-x}M_x$O$_3$ (with $M$~=~Mn, Fe or Co)\cite{sample}, these
observations cannot explain the occurrence of the MIT in the ground
state across a finite value of $x$.

For the near-ground state low-energy scale electronic structure of
these systems, we have obtained point-contact conductance spectra of
the series LaNi$_{1-x}$Mn$_x$O$_3$ at 4.2~K~(Fig.~2). 
Since the point contact tunneling conductance, to a first order, is given 
by the DOS at low temperatures and small bias voltages\cite{dos}, these 
spectra provide a representation of the changes in the
DOS with high resolution.  The conductance of
the metallic systems with $x \leq 0.1$ as a function of the applied
voltage,~$V$, clearly shows the typical cusp at $E_F$ for a
disordered system. A linear 
plot of the conductance {\it vs.} $V^{1/2}$ (inset I of
Fig.~2) establishes the characterisitic square-root dependence.
Moreover, it is known that a cusped DOS at
$E_F$ leads to a $T^{1/2}$ dependence of the conductivity at low
temperatures\cite{TVR}. Such a $T^{1/2}$ dependence has indeed
been observed\cite{our} in these samples. 
This is in sharp contrast to the predictions based on the dynamical
mean-field theory of the Hubbard-like model which predicts the
formation of an increasingly narrow Kondo-like peak at $E_F$. 
These results on the spectral
evolution with increasing disorder
 is directly relevant not only for the resistivity, 
but also for other physical
properties such as specific heat, which depend on the DOS
at $E_F$. 
Thus, while the
dynamical mean-field theory appears to describe the large-energy
scale behavior reasonably well, it fails to depict
the actual changes in the low energy scale near $E_F$.
For
the semiconducting composition ($x$~=~0.2), the conductance is {\it
linear} in~$V$ and shows the signature of a hard gap
($\approx 20$~meV) at $E_F$.
 Qualitatively similar tunneling spectra have
also been obtained\cite{thesis1} for LaNi$_{1-x}$Co$_x$O$_3$. 
However, the
transport measurements at higher temperatures for the insulating
composition\cite{our} do not reveal an activated behavior
characteristic of a hard gap. Instead, variable range hopping 
is observed even at 10 K suggesting the existence of finite localised
DOS at~$E_F$. Interestingly, tunneling conductance measured at 10 K
for $x$ = 0.2 sample exhibits a wiping out of the hard gap with a significant 
increase in the zero bias conductance as shown in the inset II.

In order to probe
the nature of single particle spectral weights in the vicinity of
$E_F$ at finite temperatures, we have studied the photoemission
spectra of LaNi$_{1-x}$Mn$_x$O$_3$ and LaNi$_{1-x}$Fe$_x$O$_3$ series
at 100~K, with $h \nu=$~55~eV shown in Fig.~3 over a narrow energy scale. 
The spectra of LaMnO$_3$~\cite{Mn} and LaFeO$_3$~\cite{Fe},
being wide band gap ($\approx$ 1.3
and 2.0~eV respectively) materials, have no appreciable weight over
the narrow energy range near $E_F$ and the spectral
function at the intermediate compositions is dominated by the
LaNiO$_3$ component of the solid solution.  The results show that
there is a systematic depletion of spectral weight within $\approx$
200~meV of $E_F$ with increasing Mn and Fe concentrations as is
clearly evident in the insets. The spectral function modifications in
the two series are very similar, indicating that the nature of
changes in the spectral functions arises only from the disorder caused
by the substitution, and follows the same  qualitative behavior. In
order to quantify the changes in the spectral function near $E_F$, we
have first analysed the LaNiO$_3$ spectrum in terms of a polynomial
DOS varying smoothly across $E_F$, broadened by the resolution
function and temperature (Fermi-Dirac statistics). The
resulting fit is shown by solid lines through the data points in Fig.~3.
Subsequently, we simulate the spectra of the metallic compositions
with $x>0.0$ by a DOS that is identical to that obtained from
LaNiO$_3$ farther away from $E_F$ as suggested by the inset, but with
a $|E-E_F|^{1/2}$ cusp near $E_F$ prompted by the depletion of
spectral weights near $E_F$ in the photoemission spectra and the
results in Fig.~2.  We find that the simulated spectra (solid 
lines) describe the
experimental results well. However, our attempt to describe the
spectra of the insulating compositions in terms of a linear DOS near
$E_F$ with a hard gap as suggested by the conductance spectra (Fig.~2)
was not successful, as shown by dashed lines in Fig.~3.
Clearly there is a lot more intensity at $E_F$ in the experimental
curve to be compatible with a gap even after taking into account the 
resolution and thermal broadenings, suggesting a finite
density of states at $E_F$ at elevated temperatures 
for the compositions with an
insulating ground state (for Mn, $x$~=~0.2, and for Fe, $x$~=~0.3 and
0.4). Thus, we attempt to describe the spectral features for these 
compositions with a DOS identical to LaNiO$_3$ in the wider energy
scale with a $|E-E_F|^m$ dependence close to $E_F$ in order to
represent the pronounced depletion of states there. We find that
$m$=1/2 provides a good description also for insulating
compositions as shown by the solid lines in the figures. We show the
extracted DOS for all the cases without the resolution and thermal 
broadening in
the insets {\scriptsize II} in Figs.~3a and b. These plots clearly
show the formation of the cusp for the substituted
compositions with a progressive depletion of states near $E_F$. At
finite temperatures, the insulating compositions appear to be
characterised by a deeper cusp existing over 
a wider energy scale.
In every case, the extracted DOS farther away from
$E_F$ is very similar to that of LaNiO$_3$. 

To summarize the present results {\it vis-a-vis} existing theories,
we find that the ground state electronic structure within the
metallic phase in the substituted compounds 
over a narrow energy scale is in agreement with
AA theory\cite{AA} with a characteristic cusp at $E_F$.
This is in contrast with the formation of a sharp resonance peak at
$E_F$ expected on the basis of Hubbard-like models\cite{EPL,priv};
we further point out that the expected resonance peak is absent even 
in the parent stoichiometric LaNiO$_3$.  
On the other hand, the formation
of a hard gap in the ground state on the insulating side of the MIT
is not consistent with the soft Coulomb gap predicted for
disordered insulating systems in presence of long range interactions
and may indicate the dominance of on-site interactions as in the
Hubbard model. It is interesting to note that the spectral
evolution over a wide energy scale (ignoring the subtle changes close
to $E_F$) is consistent with the prediction of the Hubbard-like
model with diagonal disorder. In any case, it is clear that none of
the existing theories is sufficient to describe entirely 
the evolution of the
ground state electronic structure across the MIT in substituted TMO.
The present results provide evidence of a strong influence of
disorder on the electronic structure of substituted TMO. It also
explains the systematic change in the value ($x_c$) of
critical substitution required for driving the MIT in this family of
compounds, LaNi$_{1-x}M_x$O$_3$ (with $M$ = Mn, Fe or Co). As the
substitutional element $M$ is farther away from Ni in the Periodic
Table, the strength of the disorder as measured by the difference of
the bare energies at the Ni and the $M$ sites, increases, requiring 
less number of disordered sites to drive the transition and leading to
the decreasing trend of $x_c$ as observed. Additionally, 
the present results 
provide evidence of a wiping out of the ground state hard gap of
the insulating compositions at elevated temperatures. It is a
challenging theoretical task to provide a microscopic basis for such
interesting dynamics of the gap with temperature. Even
more fundamental is the question of the ground state metal-insulator
transition in substitutional transition metal compounds with clear
evidence for a simultaneous presence of strong interaction and
disorder effects, since it is clear that no exisitng theory is able
to explain all the experimentally observed changes in the electronic
structure across the transition. Our results emphasize the need to
include the on-site Coulomb interaction as in the
Hubbard model within a theory of disorder like Altshuler-Aronov
theory which captures the low-energy features of disorder effects.
While such a theory has not been attempted so far, we believe that
the present results will motivate efforts in this direction.

Acknowledgements : DDS thanks Forschungszentrum, J\"ulich for
hospitality during a part of this work; AC and SRK thanks the CSIR,
India for financial support. This research is funded by the Department of
Science and Technology,
Government of India.

Figure captions:

Fig.~1 BIS of LaNi$_{1-x}$Mn$_x$O$_3$(open circles). 
The synthesized spectra (solid
lines) for intermediate compostions are generated by weighted average
of the spetra of the end-members, LaNiO$_3$ and LaMnO$_3$; the
relative weight of the LaMnO$_3$ component in the synthesized spectra
is shown as a function of $x$ by the open sircles in the inset, exhibiting 
a linear dependence as expected.

Fig.~2 Point-contact tunneling conductance of LaNi$_{1-x}$Mn$_x$O$_3$.
The inset I shows the conductance as a function of $\sqrt {V}$. The
solid lines are guides to the eye. The inset II shows the comparison 
of tunneling conductance for $x$ = 0.2 sample at 4.2 and 10 K, exhibiting 
a vanishing of the hard gap and an increase in the zero bias conductance.

Fig.~3 Photoemission spectra of LaNi$_{1-x}M_x$O$_3$ ((a) $M$ = Mn
and (b) $M$ = Fe) with $h \nu =
55$~eV. Spectra are superposed in the inset I to illustrate the
systematic depletion of states close to $E_F$ with increasing $x$. The full 
lines show a 
fit to simulate the depletion in the DOS with the corresponding DOS shown
in inset II.
The dashed lines are the best fits to the spectra assuming the DOS to
be linear near $E_F$ with a hard gap of 20~meV.

\end{document}